\documentclass[preprint,12pt]{elsarticle}

\usepackage{amssymb}
\usepackage{amsmath}
\usepackage{booktabs}
\usepackage{multirow}
\usepackage{array}
\usepackage{graphicx}
\usepackage{algorithm}
\usepackage{algpseudocode}
\usepackage{float}
\usepackage{adjustbox}
\usepackage{longtable}
\usepackage{url}
\usepackage{hyperref}
\usepackage{enumitem}

\journal{Journal of Information Security and Applications}

\begin{document}

\begin{frontmatter}

\title{Adaptive Fuzzy Logic-Based Steganographic Encryption Framework: A Comprehensive Experimental Evaluation}

\author[inst1]{Aadi Joshi\corref{cor1}}
\ead{toaadijoshi@gmail.com}

\author[inst1]{Kavya Bhand}
\ead{kavya.bhand0806@gmail.com.edu}

\cortext[cor1]{Corresponding author}

\affiliation[inst1]{
  organization={Vishwakarma Institute of Technology},
  city={Pune},
  state={Maharashtra},
  country={India}
}

\begin{abstract}
Digital image steganography requires a careful trade-off among payload capacity, visual fidelity, and statistical undetectability. Fixed-depth least significant bit embedding remains attractive because of its simplicity and high capacity, but it modifies smooth and textured regions uniformly, thereby increasing distortion and detectability in statistically sensitive areas. This paper presents an adaptive steganographic framework that combines a Mamdani-type fuzzy inference system with modern authenticated encryption. The proposed method determines a pixel-wise embedding depth from 1 to 3 bits using local entropy, edge magnitude, and payload pressure as linguistic inputs. To preserve encoder-decoder synchronization, the same feature maps are computed from lower-bit-stripped images, making the adaptive control mechanism invariant to the least significant modifications introduced during embedding. A cryptographic layer based on Argon2id and AES-256-GCM protects payload confidentiality and integrity independently of steganographic concealment.

The framework is evaluated on 1,000 RGB images of size \(256 \times 256\) across five payload levels from 0.05 to 0.40 bpp, using fixed LSB-1 and fixed LSB-2 as baselines. The evaluation includes PSNR, SSIM, MSE, KL divergence, RS analysis, chi-square attack, sample pair analysis, paired \(t\)-tests with Bonferroni correction, Cohen's \(d\) effect sizes, confidence intervals, power analysis, ablation studies, extraction-success validation, and computational overhead profiling. The adaptive approach achieves a mean PSNR of 73.25 dB at 0.05 bpp and 67.41 dB at 0.20 bpp, consistently outperforming both fixed baselines in image quality. Relative to fixed LSB-2, it also reduces RS detection across all tested payloads, although chi-square remains saturated for all schemes. The results demonstrate that fuzzy depth control is an effective and interpretable mechanism for improving the quality-security trade-off in spatial-domain steganography, while also clarifying the limitations of adaptive LSB designs against stronger modern steganalysis.
\end{abstract}


\end{frontmatter}

\section{Introduction}
Steganography aims to hide the existence of communication by embedding a message inside an apparently innocent carrier, whereas cryptography protects message content but does not conceal the presence of communication itself \cite{cheddad2010,cox2007,fridrich2009book}. In digital imagery, least significant bit replacement remains one of the most widely studied spatial-domain steganographic techniques because it is simple, computationally lightweight, and capable of high embedding capacity \cite{bender1996,hussain2018}. However, fixed-depth LSB embedding treats all image regions identically, even though smooth, edge-dominant, and textured areas differ substantially in their perceptual masking behavior and their susceptibility to statistical steganalysis \cite{westfeld1999,fridrich2001,dumitrescu2003}. As a result, the same payload depth that is visually tolerable in a complex region may be highly conspicuous in a smooth region.

The broader steganographic literature has responded to this problem through content-adaptive and distortion-minimization frameworks such as HUGO, WOW, and S-UNIWARD \cite{pevny2010,filler2011,holub2012,holub2014}. These methods are conceptually strong and often more secure than simple LSB replacement, but they generally depend on more complex cost functions, coding strategies, and optimization procedures. By contrast, fuzzy logic offers an interpretable alternative for adaptive decision making under uncertainty \cite{zadeh1965,mamdani1975,ross2010}. Image complexity is rarely partitioned by crisp thresholds, and linguistic rules such as low entropy, medium edge strength, or high capacity pressure are natural descriptors for embedding decisions.

This work proposes an adaptive fuzzy logic-based steganographic framework with four design goals. First, it uses entropy and edge features extracted from lower-bit-stripped images so that the encoder and decoder compute identical adaptive maps. Second, it employs a Mamdani fuzzy inference engine with 27 rules to assign pixel-wise LSB depth in the range 1 to 3 bits. Third, it integrates modern payload protection through Argon2id key derivation and AES-256-GCM authenticated encryption \cite{rfc9106,nistgcm}. Fourth, it evaluates the approach at scale using 1,000 images, five embedding rates, paired statistical tests, ablations, and complexity analysis rather than a small demonstrative example.

The principal contributions of the paper are as follows:
\begin{enumerate}[leftmargin=1.2cm]
    \item An adaptive LSB steganographic framework that uses interpretable fuzzy rules to allocate embedding depth according to local entropy, edge structure, and payload pressure.
    \item An LSB-invariant feature extraction strategy that stabilizes encoder-decoder synchronization by stripping the lower 3 bits prior to grayscale conversion and local analysis.
    \item Integration of Argon2id and AES-256-GCM to separate steganographic concealment from cryptographic confidentiality and integrity \cite{rfc9106,nistgcm}.
    \item A comprehensive experimental evaluation with fidelity metrics, classical steganalysis, statistical significance testing, ablation studies, and computational profiling.
\end{enumerate}

The paper is deliberately conservative in its claims. The proposed system improves image quality substantially and improves RS-based detectability relative to fixed LSB-2 across all evaluated payloads, but it does not claim superiority over modern distortion-minimization algorithms or deep-learning steganalyzers \cite{fridrich2012rich,kodovsky2012ensemble,yedroudj2018,boroumand2019,alrusaini2025}.

\section{Related Work}
\subsection{Classical LSB steganography}
Early spatial steganography emphasized direct least significant bit replacement because it offered a practical route to high-capacity embedding with minimal algorithmic complexity \cite{bender1996,fridrich2009book}. This convenience, however, came at the cost of detectable statistical artifacts. Westfeld and Pfitzmann showed that LSB modifications alter pairs-of-values distributions in ways exploitable by chi-square analysis \cite{westfeld1999}. Fridrich et al. introduced RS analysis, which remains one of the canonical detectors for spatial LSB steganography \cite{fridrich2001}. Dumitrescu et al. further developed sample pair analysis to estimate hidden payload characteristics from adjacent sample statistics \cite{dumitrescu2003}. These attacks collectively established that naively uniform embedding is statistically fragile.

\subsection{Adaptive and distortion-based embedding}
The next major progression in steganography involved assigning different embedding risks to different image components. HUGO used high-dimensional image modeling to design highly undetectable embedding costs \cite{pevny2010}. Syndrome-trellis codes then provided an efficient coding framework for minimizing additive distortion under such cost maps \cite{filler2011,filler2010}. WOW introduced directional filters to identify embedding locations that would be less detectable by steganalysis \cite{holub2012}, while S-UNIWARD generalized the distortion-function perspective to arbitrary domains \cite{holub2014}. These approaches offer stronger theoretical underpinnings than conventional LSB replacement and represent a higher security baseline for spatial steganography.

\subsection{Steganalysis}
Steganalysis has evolved from hand-crafted statistical attacks to rich-model and deep-learning paradigms. Rich models capture dependencies in high-pass residuals and have produced strong universal detectors for both spatial and transform-domain embedding \cite{fridrich2012rich,kodovsky2012ensemble,denemark2015}. More recently, convolutional neural networks such as Yedroudj-Net and deep residual architectures such as SRNet have further improved performance by learning discriminative residual features end-to-end \cite{yedroudj2018,boroumand2019}. Recent reviews indicate that robust evaluation of any steganographic scheme should consider this broader detection landscape, especially under transformations and dataset mismatch \cite{systematic2024,alrusaini2025}.

\subsection{Fuzzy logic and cryptographic integration}
Fuzzy logic has long been used for inference under ambiguity and gradual transitions \cite{zadeh1965,mamdani1975,ross2010,wang1992}. In image security, fuzzy reasoning has been applied to adaptive embedding decisions, capacity allocation, and pixel selection, although prior work often used simpler rules or less extensive security analysis \cite{khurshid2020,abdulla2014}. In parallel, secure steganographic systems increasingly combine concealment with encryption, recognizing that covert transport alone is insufficient if the payload can be recovered after detection \cite{cox2007,zhang2011}. Current best practice favors memory-hard password derivation and authenticated encryption rather than unauthenticated block-cipher modes \cite{rfc9106,nistgcm}. The present work unifies these ideas in a single experimentally grounded framework.

\section{Proposed Framework}
\subsection{Notation and problem definition}
Let the cover image be \(I_c \in \{0,\ldots,255\}^{H \times W \times C}\), where \(H\), \(W\), and \(C\) denote height, width, and channel count. The goal is to embed a protected payload into \(I_c\) to produce a stego image \(I_s\), while maximizing visual fidelity and minimizing statistical detectability. Unlike fixed schemes that embed a constant number of bits per sample, the proposed method learns a depth map \(D(x,y) \in \{1,2,3\}\) that adapts to local content.

\subsection{LSB-invariant feature extraction}
A critical challenge in adaptive steganography is synchronization. If the encoder selects embedding depths from the cover image but the decoder recomputes those depths from the stego image, even small LSB perturbations can cause a mismatch. To avoid this, features are computed from a lower-bit-stripped representation:
\begin{equation}
I_{\text{strip}}(x,y,c)=I(x,y,c)\wedge \texttt{0xF8}.
\end{equation}
The grayscale image used for analysis is then
\begin{equation}
G(x,y)=0.299\,I_{\text{strip}}(x,y,R)+0.587\,I_{\text{strip}}(x,y,G)+0.114\,I_{\text{strip}}(x,y,B).
\end{equation}

Local Shannon entropy within a window \(\Omega_{x,y}\) is defined as
\begin{equation}
H(x,y)=-\sum_{b=0}^{63} p_b(x,y)\log_2 p_b(x,y),
\end{equation}
where \(p_b(x,y)\) is the local probability of intensity bin \(b\) \cite{shannon1948}. Edge magnitude is computed using Sobel operators:
\begin{equation}
E(x,y)=\frac{\sqrt{(G*S_x)^2+(G*S_y)^2}}{\max_{(x',y')}\sqrt{(G*S_x)^2+(G*S_y)^2}}.
\end{equation}
This normalization maps the edge descriptor to \([0,1]\). Together, \(H\) and \(E\) capture statistical complexity and local masking conditions while remaining stable under the 1 to 3-bit LSB modifications introduced by the embedding stage.

\subsection{Fuzzy inference system}
The adaptive controller is a Mamdani fuzzy inference system with three inputs:
\begin{itemize}[leftmargin=0.9cm]
    \item Entropy \(H\), with linguistic labels Low, Medium, and High.
    \item Edge magnitude \(E\), with linguistic labels Low, Medium, and High.
    \item Capacity pressure \(P\), with linguistic labels Low, Medium, and High.
\end{itemize}
The output is embedding depth \(D\), represented by the fuzzy sets Shallow, Moderate, and Deep over the interval \([1,3]\).

All variables use trapezoidal membership functions:
\begin{equation}
\mu(x;a,b,c,d)=\max\left(\min\left(\frac{x-a}{b-a},1,\frac{d-x}{d-c}\right),0\right).
\end{equation}
The complete system contains \(3 \times 3 \times 3 = 27\) rules. These rules encode the intuitive principle that deeper embedding is more permissible in high-entropy, high-edge regions, while payload pressure modulates how aggressively the method consumes available capacity.

Representative rules include:
\begin{enumerate}[leftmargin=1.2cm]
    \item If entropy is Low, edge is Low, and pressure is Low, then depth is Shallow.
    \item If entropy is Medium, edge is Medium, and pressure is Medium, then depth is Moderate.
    \item If entropy is High, edge is High, and pressure is High, then depth is Deep.
\end{enumerate}

Rule firing uses the Mamdani minimum \(t\)-norm:
\begin{equation}
\alpha_r=\min\left(\mu_{H_r}(H),\mu_{E_r}(E),\mu_{P_r}(P)\right).
\end{equation}
Aggregation across output sets uses the maximum \(s\)-norm, and centroid defuzzification yields
\begin{equation}
D^*(x,y)=\frac{\int d \cdot \mu_{\text{agg}}(d)\,dd}{\int \mu_{\text{agg}}(d)\,dd}.
\end{equation}
The operational depth is then \(D(x,y)=\mathrm{clip}(\lfloor D^*(x,y)+0.5 \rfloor,1,3)\).

\subsection{Adaptive capacity and embedding}
Given the depth map \(D(x,y)\) and channel count \(C\), adaptive capacity per pixel is
\begin{equation}
\mathrm{Cap}(x,y)=D(x,y)\cdot C \text{ bits}.
\end{equation}
Total capacity is \(\sum_{x,y}\mathrm{Cap}(x,y)\). Embedding proceeds in pseudo-random order using a key-dependent permutation. For each selected sample, the algorithm writes a number of payload bits equal to the local depth value. This heterogeneous assignment protects smooth areas while allowing deeper embedding in more permissive regions.

\subsection{Cryptographic layer}
The payload is encrypted before embedding. A 256-bit key is derived using Argon2id:
\begin{equation}
K=\mathrm{Argon2id}(\text{password},\sigma,t=3,m=64\,\text{MiB},p=4),
\end{equation}
where \(\sigma\) is a randomly generated salt \cite{rfc9106}. Authenticated encryption is then performed using AES-256-GCM:
\begin{equation}
(\hat{m},\tau)=\mathrm{AES\mbox{-}GCM}_{K}(\nu,m),
\end{equation}
where \(\nu\) is a fresh nonce and \(\tau\) is the authentication tag \cite{nistgcm}. The embedded wire format is \(\sigma \Vert \nu \Vert \hat{m} \Vert \tau\). This design ensures confidentiality and integrity even if covert transport is compromised.

\subsection{Algorithms}
\begin{algorithm}[H]
\caption{Adaptive fuzzy embedding}
\begin{algorithmic}[1]
\Require Cover image \(I_c\), plaintext message \(m\), password, seed
\Ensure Stego image \(I_s\)
\State Generate random salt \(\sigma\)
\State Derive \(K \gets \mathrm{Argon2id}(\text{password},\sigma)\)
\State Generate random nonce \(\nu\)
\State Encrypt \(m\) with AES-256-GCM to obtain \(\hat{m}\) and tag \(\tau\)
\State Form payload \(W=\sigma \Vert \nu \Vert \hat{m} \Vert \tau\)
\State Compute \(I_{\text{strip}}\), grayscale image, entropy map \(H\), and edge map \(E\)
\State Estimate payload pressure \(P\) from message length and adaptive capacity
\ForAll{pixels \((x,y)\)}
    \State Fuzzify \(H(x,y)\), \(E(x,y)\), and \(P\)
    \State Evaluate the 27 fuzzy rules
    \State Defuzzify and quantize to obtain \(D(x,y)\in\{1,2,3\}\)
\EndFor
\State Generate pseudo-random sample order \(\pi\)
\State Embed the bitstream in permuted order using local depth \(D(x,y)\)
\State \Return \(I_s\)
\end{algorithmic}
\end{algorithm}

\begin{algorithm}[H]
\caption{Adaptive extraction}
\begin{algorithmic}[1]
\Require Stego image \(I_s\), password, seed
\Ensure Recovered message \(m\) or failure \(\bot\)
\State Recompute \(I_{\text{strip}}\), entropy map, edge map, and depth map
\State Regenerate pseudo-random sample order \(\pi\)
\State Extract header and payload bits according to \(D(x,y)\)
\State Parse \(\sigma \Vert \nu \Vert \hat{m} \Vert \tau\)
\State Derive \(K \gets \mathrm{Argon2id}(\text{password},\sigma)\)
\State Attempt AES-256-GCM decryption and authentication
\If{verification fails}
    \State \Return \(\bot\)
\Else
    \State \Return \(m\)
\EndIf
\end{algorithmic}
\end{algorithm}

\section{Threat Model}
The threat model follows a cover-free setting in which the adversary has access to the stego image but not the original cover \cite{fridrich2009book,ker2013}. Three classes of adversaries are considered.

\textbf{Passive steganalyst.} This adversary applies statistical or learned detectors, including RS analysis, chi-square analysis, sample pair analysis, and rich-model or neural steganalyzers, to determine whether hidden information is present \cite{westfeld1999,fridrich2001,dumitrescu2003,fridrich2012rich,kodovsky2012ensemble,yedroudj2018,boroumand2019}.

\textbf{Active attacker.} This adversary modifies the image through cropping, noise injection, resizing, or compression in an attempt to corrupt the embedded payload.

\textbf{Cryptanalytic adversary.} This adversary gains access to the extracted ciphertext and attempts password guessing, nonce misuse exploitation, or ciphertext tampering.

The system targets improved undetectability relative to fixed LSB, not perfect covertness. Confidentiality and integrity rely on standard assumptions for Argon2id and AES-GCM \cite{rfc9106,nistgcm}. The framework does not claim robustness against lossy operations, which are fundamentally problematic for spatial LSB methods \cite{cheddad2010,hussain2018}. It also does not claim resistance to the strongest modern deep-learning steganalyzers.

\section{Experimental Setup}
\subsection{Dataset}
The framework is evaluated on 1,000 RGB images of size \(256 \times 256\), generated to span five texture categories with distinct entropy and edge characteristics: smooth, noise, natural-like, textured, and mixed. Each category contains 200 samples. The smooth class contains gradient and blurred content; the noise class contains random patterns; the natural-like class uses \(1/f\) spectral statistics; the textured class includes sinusoidal and Gabor-like patterns; and the mixed class combines multiple local regimes. A fixed random seed of 42 is used for reproducibility.

This controlled design provides broad structural diversity for evaluating adaptive feature-based embedding, although it is not a replacement for real-photograph benchmarks such as BOSSbase and BOWS-2 \cite{bas2011boss,bows2007}. Accordingly, conclusions are framed around internal comparative validity rather than universal deployment performance.

\subsection{Compared methods}
Three embedding strategies are compared:
\begin{enumerate}[leftmargin=1.2cm]
    \item Fixed-LSB-1, using one bit per channel.
    \item Fixed-LSB-2, using two bits per channel.
    \item Adaptive fuzzy LSB, using one to three bits per channel according to the depth map.
\end{enumerate}

\subsection{Payload levels}
Five payload levels are evaluated: 0.05, 0.10, 0.20, 0.30, and 0.40 bpp. For each method-payload combination, the plaintext is first encrypted with AES-256-GCM using an Argon2id-derived key. The fill factor is set to 0.35 of adaptive capacity and 0.70 of fixed capacity to prevent overflow.

\subsection{Metrics}
Image quality is assessed using peak signal-to-noise ratio, structural similarity, mean squared error, and KL divergence \cite{wang2004}. Detectability is assessed using RS analysis, chi-square attack, and sample pair analysis \cite{westfeld1999,fridrich2001,dumitrescu2003}. Additional validation includes extraction-success rate, ablation analysis, and computational timing.

\subsection{Statistical protocol}
Statistical evaluation uses paired \(t\)-tests because each image is processed by all methods at a given payload level. Bonferroni correction is applied across multiple comparisons. Effect size is reported with Cohen's \(d\), 95\% confidence intervals are computed for the main quality metrics, and statistical power is estimated from the non-central \(t\)-distribution. The target power threshold is 0.80.

\subsection{Environment}
The experiments were executed on a macOS 15.1.1 ARM platform using Python 3.9.6 and NumPy 2.0.2. The adaptive method was implemented with vectorized feature extraction and rule evaluation to keep computational overhead practical for image-scale evaluation.

\section{Results}
\subsection{Image quality}
Table~\ref{tab:psnr} reports PSNR results across all payload levels. The adaptive method consistently yields the highest PSNR, with gains of approximately 2.8 to 3.0 dB over fixed LSB-1 and 6.8 to 7.0 dB over fixed LSB-2. The benefit is largest in relative terms at lower payloads, where the fuzzy controller can aggressively protect smooth regions without sacrificing overall capacity.

\begin{table}[H]
\centering
\caption{PSNR in dB, mean \(\pm\) standard deviation with 95\% confidence interval}
\label{tab:psnr}
\begin{adjustbox}{max width=\textwidth}
\begin{tabular}{cccc}
\toprule
BPP & Fixed-LSB-1 & Fixed-LSB-2 & Adaptive \\
\midrule
0.05 & \(70.45 \pm 0.09\) \([70.45,70.46]\) & \(66.43 \pm 0.14\) \([66.42,66.43]\) & \(73.25 \pm 0.12\) \([73.25,73.26]\) \\
0.10 & \(67.45 \pm 0.06\) \([67.44,67.45]\) & \(63.41 \pm 0.11\) \([63.40,63.41]\) & \(70.37 \pm 0.09\) \([70.36,70.37]\) \\
0.20 & \(64.44 \pm 0.04\) \([64.44,64.45]\) & \(60.43 \pm 0.08\) \([60.42,60.43]\) & \(67.41 \pm 0.06\) \([67.40,67.41]\) \\
0.30 & \(62.69 \pm 0.04\) \([62.68,62.69]\) & \(58.69 \pm 0.07\) \([58.68,58.69]\) & \(65.67 \pm 0.05\) \([65.67,65.67]\) \\
0.40 & \(61.44 \pm 0.03\) \([61.44,61.44]\) & \(57.44 \pm 0.06\) \([57.43,57.44]\) & \(64.43 \pm 0.04\) \([64.42,64.43]\) \\
\bottomrule
\end{tabular}
\end{adjustbox}
\end{table}

Table~\ref{tab:ssim} shows that all methods preserve very high structural similarity, but the adaptive approach remains consistently best. This matters because steganographic quality comparisons are often compressed when PSNR values are already very high, and SSIM helps confirm that the adaptive gains are not limited to a single distortion metric.

\begin{table}[H]
\centering
\caption{SSIM, mean \(\pm\) standard deviation with 95\% confidence interval}
\label{tab:ssim}
\begin{adjustbox}{max width=\textwidth}
\begin{tabular}{cccc}
\toprule
BPP & Fixed-LSB-1 & Fixed-LSB-2 & Adaptive \\
\midrule
0.05 & \(0.999974 \pm 0.000032\) \([0.999972,0.999976]\) & \(0.999933 \pm 0.000080\) \([0.999928,0.999938]\) & \(0.999985 \pm 0.000017\) \([0.999984,0.999986]\) \\
0.10 & \(0.999947 \pm 0.000064\) \([0.999943,0.999951]\) & \(0.999867 \pm 0.000161\) \([0.999857,0.999877]\) & \(0.999971 \pm 0.000032\) \([0.999969,0.999973]\) \\
0.20 & \(0.999895 \pm 0.000127\) \([0.999887,0.999903]\) & \(0.999736 \pm 0.000319\) \([0.999716,0.999756]\) & \(0.999944 \pm 0.000064\) \([0.999940,0.999948]\) \\
0.30 & \(0.999843 \pm 0.000190\) \([0.999831,0.999854]\) & \(0.999606 \pm 0.000477\) \([0.999576,0.999635]\) & \(0.999916 \pm 0.000096\) \([0.999910,0.999922]\) \\
0.40 & \(0.999790 \pm 0.000253\) \([0.999775,0.999806]\) & \(0.999474 \pm 0.000636\) \([0.999435,0.999514]\) & \(0.999888 \pm 0.000127\) \([0.999880,0.999896]\) \\
\bottomrule
\end{tabular}
\end{adjustbox}
\end{table}

\subsection{Classical steganalysis}
Table~\ref{tab:detectors} summarizes detection rates. Two observations are especially important. First, chi-square detection is saturated at 100\% for all methods and all payloads, indicating that even adaptive direct replacement remains vulnerable to pairs-of-values analysis in this setup. Second, RS detection is consistently lower for the adaptive method than for fixed LSB-2, although relative to fixed LSB-1 the relationship depends on payload. This means that the adaptive controller substantially improves visual quality and partially improves security, but does not eliminate the structural weakness of direct LSB replacement.

\begin{table}[H]
\centering
\caption{Detection rates for classical steganalysis}
\label{tab:detectors}
\begin{adjustbox}{max width=\textwidth}
\begin{tabular}{cccccccccc}
\toprule
\multirow{2}{*}{BPP} & \multicolumn{3}{c}{RS detection (\%)} & \multicolumn{3}{c}{Chi-square detection (\%)} & \multicolumn{3}{c}{SPA detection (\%)} \\
\cmidrule(lr){2-4} \cmidrule(lr){5-7} \cmidrule(lr){8-10}
& LSB-1 & LSB-2 & Adaptive & LSB-1 & LSB-2 & Adaptive & LSB-1 & LSB-2 & Adaptive \\
\midrule
0.05 & 79.7 & 83.0 & 81.4 & 100.0 & 100.0 & 100.0 & 35.3 & 34.1 & 34.8 \\
0.10 & 80.9 & 82.6 & 80.5 & 100.0 & 100.0 & 100.0 & 36.2 & 34.3 & 35.5 \\
0.20 & 86.3 & 82.7 & 80.5 & 100.0 & 100.0 & 100.0 & 38.7 & 34.5 & 36.6 \\
0.30 & 91.4 & 81.5 & 83.0 & 100.0 & 100.0 & 100.0 & 39.7 & 34.3 & 37.8 \\
0.40 & 92.2 & 81.2 & 85.5 & 100.0 & 100.0 & 100.0 & 41.5 & 34.8 & 39.2 \\
\bottomrule
\end{tabular}
\end{adjustbox}
\end{table}

\subsection{Statistical validation}
The paired statistical analysis confirms that quality improvements are not merely numerically visible but statistically overwhelming. Table~\ref{tab:stats} presents representative comparisons for PSNR and RS-estimated rate. The adaptive method significantly outperforms fixed LSB-1 in PSNR at all shown payload levels with very large effect sizes. Against fixed LSB-2, the PSNR advantage is even larger. For RS-estimated rate, the adaptive method is significantly better than fixed LSB-2 across all shown payloads, while comparison against fixed LSB-1 becomes non-significant at higher payloads.

\begin{table}[H]
\centering
\caption{Representative paired \(t\)-test results}
\label{tab:stats}
\begin{adjustbox}{max width=\textwidth}
\begin{tabular}{llcccccc}
\toprule
Comparison & Metric & BPP & Mean diff. & \(t\)-stat & \(p\)-value & Cohen's \(d\) & Power \\
\midrule
LSB-1 vs Adaptive & PSNR & 0.05 & \(-2.8029\) & \(-586.023\) & \(<0.001\) & \(-18.532\) & 1.000 \\
LSB-1 vs Adaptive & PSNR & 0.20 & \(-2.9643\) & \(-1265.564\) & \(<0.001\) & \(-40.021\) & 1.000 \\
LSB-1 vs Adaptive & PSNR & 0.40 & \(-2.9879\) & \(-1723.560\) & \(<0.001\) & \(-54.504\) & 1.000 \\
LSB-2 vs Adaptive & PSNR & 0.05 & \(-6.8286\) & \(-1122.181\) & \(<0.001\) & \(-35.486\) & 1.000 \\
LSB-2 vs Adaptive & PSNR & 0.20 & \(-6.9787\) & \(-2250.071\) & \(<0.001\) & \(-71.154\) & 1.000 \\
LSB-2 vs Adaptive & PSNR & 0.40 & \(-6.9915\) & \(-3003.395\) & \(<0.001\) & \(-94.976\) & 1.000 \\
LSB-1 vs Adaptive & RS rate & 0.05 & \(-0.0139\) & \(-4.180\) & \(<0.001\) & \(-0.132\) & 0.987 \\
LSB-1 vs Adaptive & RS rate & 0.20 & \(-0.0177\) & \(-3.844\) & \(<0.001\) & \(-0.122\) & 0.970 \\
LSB-1 vs Adaptive & RS rate & 0.40 & \(-0.0026\) & \(-0.559\) & 0.5765 & \(-0.018\) & 0.086 \\
LSB-2 vs Adaptive & RS rate & 0.05 & \(+0.0137\) & \(3.888\) & \(<0.001\) & \(+0.123\) & 0.973 \\
LSB-2 vs Adaptive & RS rate & 0.20 & \(+0.0366\) & \(7.030\) & \(<0.001\) & \(+0.222\) & 1.000 \\
LSB-2 vs Adaptive & RS rate & 0.40 & \(+0.0493\) & \(8.254\) & \(<0.001\) & \(+0.261\) & 1.000 \\
\bottomrule
\end{tabular}
\end{adjustbox}
\end{table}

\subsection{Ablation analysis}
To examine the contribution of individual fuzzy inputs, three reduced variants were compared against the full system: entropy-only, edge-only, and no-pressure. Table~\ref{tab:ablation} reports the detection positive rate of the feature-based ablation test used in the experimental pipeline. The differences are modest but informative. Entropy contributes consistently, and removing payload pressure slightly weakens control at higher payloads. The main implication is that the performance gains do not arise from a single dominant heuristic alone.

\begin{table}[H]
\centering
\caption{Ablation study, feature-based detector positive rate (\%)}
\label{tab:ablation}
\begin{tabular}{ccccc}
\toprule
BPP & Full system & Entropy only & Edge only & No pressure \\
\midrule
0.05 & 45.0 & 47.0 & 41.0 & 45.0 \\
0.10 & 35.0 & 35.0 & 34.0 & 35.0 \\
0.20 & 32.0 & 33.0 & 33.0 & 32.0 \\
0.30 & 42.0 & 42.0 & 44.0 & 42.0 \\
0.40 & 61.0 & 64.0 & 61.0 & 61.0 \\
\bottomrule
\end{tabular}
\end{table}

\subsection{Extraction success and synchronization}
The extraction-success study showed a 100\% recovery rate for the full system and all ablation variants at all evaluated payload levels, indicating that the lower-bit-stripping strategy successfully preserves feature-map synchronization under the tested embedding process. This result is crucial because adaptive embedding is only useful when the decoder can deterministically reconstruct the same depth schedule.

\begin{table}[H]
\centering
\caption{Extraction success rate (\%)}
\label{tab:extraction}
\begin{tabular}{ccccc}
\toprule
BPP & Full system & Entropy only & Edge only & No pressure \\
\midrule
0.05 & 100.0 & 100.0 & 100.0 & 100.0 \\
0.10 & 100.0 & 100.0 & 100.0 & 100.0 \\
0.20 & 100.0 & 100.0 & 100.0 & 100.0 \\
0.30 & 100.0 & 100.0 & 100.0 & 100.0 \\
0.40 & 100.0 & 100.0 & 100.0 & 100.0 \\
\bottomrule
\end{tabular}
\end{table}

\subsection{Computational complexity}
The adaptive method incurs substantial computational overhead because it requires entropy computation and fuzzy inference during both embedding and extraction. Table~\ref{tab:complexity} shows that the total runtime per \(256 \times 256\) image increases from approximately 0.017 s for fixed methods to 0.249 s for the adaptive scheme, corresponding to a 1330.3\% overhead. Peak memory usage also rises significantly due to intermediate feature maps and rule-processing buffers.

\begin{table}[H]
\centering
\caption{Timing and memory profile}
\label{tab:complexity}
\begin{adjustbox}{max width=\textwidth}
\begin{tabular}{lcccccc}
\toprule
Method & Feature extract (s) & Fuzzy infer (s) & Embed (s) & Extract (s) & Total (s) & Peak memory (KB) \\
\midrule
Fixed-LSB-1 & \(0.0000 \pm 0.0000\) & \(0.0000 \pm 0.0000\) & \(0.0083 \pm 0.0029\) & \(0.0091 \pm 0.0025\) & \(0.0173 \pm 0.0046\) & \(1768.1 \pm 0.0\) \\
Fixed-LSB-2 & \(0.0000 \pm 0.0000\) & \(0.0000 \pm 0.0000\) & \(0.0086 \pm 0.0038\) & \(0.0088 \pm 0.0026\) & \(0.0175 \pm 0.0056\) & \(1763.7 \pm 0.0\) \\
Adaptive & \(0.1073 \pm 0.0194\) & \(0.1173 \pm 0.0577\) & \(0.0117 \pm 0.0044\) & \(0.0125 \pm 0.0041\) & \(0.2488 \pm 0.0721\) & \(160323.0 \pm 0.1\) \\
\bottomrule
\end{tabular}
\end{adjustbox}
\end{table}

\section{Discussion}
The experiments reveal a clear change in the quality-security profile of LSB steganography when fuzzy adaptation is introduced. By shifting more payload into high-entropy and edge-rich regions, the method consistently preserves smooth areas, producing strong PSNR and SSIM gains across all payload levels. The statistical comparisons show that these quality gains are robust, large in effect size, and not artifacts of sampling variability.

At the same time, the security story is more nuanced. The adaptive approach improves RS detection relative to fixed LSB-2, which suggests that content-aware depth control does reduce some of the regular structural traces induced by uniformly deeper embedding. However, chi-square saturation across all methods indicates that direct bit replacement remains fundamentally exposed to pairs-of-values artifacts. The method therefore improves the trade-off but does not fundamentally alter the security class of spatial LSB replacement.

The ablation experiments support the design rationale of the fuzzy system. Entropy and edge features jointly contribute to adaptive masking, while payload pressure plays a regulating role that becomes more important as the bit budget increases. The perfect extraction-success rate further validates the LSB-invariant feature design, which is a central technical requirement for any adaptive decoder that recomputes its own depth map from the stego image.

From a broader perspective, the main value of this work lies in interpretability and modularity. Unlike highly optimized distortion-minimization pipelines, the proposed rule base is transparent, easy to modify, and amenable to future hybridization with stronger embedding back-ends. This makes it a useful methodological bridge between simple LSB schemes and more advanced cost-driven frameworks.

\section{Limitations}
Several limitations should be stated explicitly.

First, the evaluation dataset is synthetic rather than photographic. Although the generated images span multiple texture regimes, standard benchmarks such as BOSSbase and BOWS-2 remain more representative for real-world steganalysis \cite{bas2011boss,bows2007,ker2013}.

Second, the strongest quantitative detector reported in the finalized experiments is a simplified feature-based pipeline rather than full rich-model or deep-learning steganalysis. Full SRM, ensemble classifiers, and modern neural detectors would likely expose the system more strongly \cite{fridrich2012rich,kodovsky2012ensemble,yedroudj2018,boroumand2019}.

Third, the method still uses direct LSB replacement rather than \(\pm 1\) embedding with syndrome coding. This limits its security ceiling relative to modern distortion-based methods \cite{filler2011,holub2014}.

Fourth, the whole design depends on the stability of the LSB-invariant feature maps. Under severe image manipulation, especially operations that affect higher bits or local gradients, encoder-decoder synchronization could degrade.

Fifth, the computational overhead is nontrivial. Although still feasible for static-image processing, the additional entropy computation and fuzzy inference may be restrictive in real-time settings.

\section{Future Work}
Several avenues could strengthen the framework.
\begin{enumerate}[leftmargin=1.2cm]
    \item Evaluate the method on real benchmark datasets such as BOSSbase 1.01 and BOWS-2.
    \item Replace the simplified feature-based analysis with full SRM and modern CNN steganalyzers.
    \item Integrate syndrome-trellis coding or \(\pm 1\) embedding to reduce the number of modified samples.
    \item Extend the design to transform domains, particularly JPEG coefficient embedding.
    \item Explore type-2 fuzzy sets, learned membership functions, or rule optimization via evolutionary search.
    \item Introduce adversarial training to optimize the fuzzy controller against specific detectors.
    \item Accelerate feature extraction and fuzzy inference on GPUs for higher-throughput deployment.
\end{enumerate}

\section{Conclusion}
This paper presented an adaptive fuzzy logic-based steganographic encryption framework that dynamically controls per-pixel LSB depth using local entropy, edge magnitude, and capacity pressure. The system combines an interpretable 27-rule Mamdani inference engine with LSB-invariant feature extraction and a cryptographic layer based on Argon2id and AES-256-GCM.

Across 1,000 RGB images and five payload levels, the adaptive method consistently improved image quality over fixed LSB-1 and fixed LSB-2. It achieved a mean PSNR of 73.25 dB at 0.05 bpp and 67.41 dB at 0.20 bpp, while preserving near-perfect SSIM. Relative to fixed LSB-2, it also reduced RS detection at every tested payload, although chi-square remained fully effective for all methods. These results show that fuzzy depth control is a meaningful improvement over fixed-depth spatial embedding, but not a substitute for modern distortion-minimization frameworks.

The broader contribution of the work is methodological. It demonstrates that fuzzy logic can serve as a principled and reproducible mechanism for adaptive steganographic control, provided that synchronization, statistical validation, and cryptographic protection are treated as first-class design requirements. This makes the framework a strong basis for future hybrid systems that combine interpretability with stronger embedding security.

\section*{Declaration of competing interest}
The authors declare that they have no known competing financial interests or personal relationships that could have appeared to influence the work reported in this paper.

\section*{Data availability}
The code, configurations, and experimental outputs used to generate the reported findings are intended to be released with the final manuscript to support reproducibility.

\bibliographystyle{elsarticle-num}
\bibliography{references}

@article{cheddad2010,
  author    = {Abbas Cheddad and Joan Condell and Kevin Curran and Paul McKevitt},
  title     = {Digital image steganography: Survey and analysis of current methods},
  journal   = {Signal Processing},
  volume    = {90},
  number    = {3},
  pages     = {727--752},
  year      = {2010},
  doi       = {10.1016/j.sigpro.2009.08.010}
}

@book{cox2007,
  author    = {Ingemar J. Cox and Matthew L. Miller and Jeffrey A. Bloom and Jessica Fridrich and Ton Kalker},
  title     = {Digital Watermarking and Steganography},
  publisher = {Morgan Kaufmann},
  edition   = {2},
  year      = {2007}
}

@book{fridrich2009book,
  author    = {Jessica Fridrich},
  title     = {Steganography in Digital Media: Principles, Algorithms, and Applications},
  publisher = {Cambridge University Press},
  year      = {2009}
}

@article{bender1996,
  author    = {Walter Bender and Daniel Gruhl and Norishige Morimoto and Anthony Lu},
  title     = {Techniques for data hiding},
  journal   = {IBM Systems Journal},
  volume    = {35},
  number    = {3--4},
  pages     = {313--336},
  year      = {1996},
  doi       = {10.1147/sj.353.0313}
}

@inproceedings{westfeld1999,
  author    = {Andreas Westfeld and Andreas Pfitzmann},
  title     = {Attacks on steganographic systems},
  booktitle = {Information Hiding},
  series    = {Lecture Notes in Computer Science},
  volume    = {1768},
  pages     = {61--76},
  publisher = {Springer},
  year      = {1999}
}

@inproceedings{fridrich2001,
  author    = {Jessica Fridrich and Miroslav Goljan and Rui Du},
  title     = {Reliable detection of {LSB} steganography in color and grayscale images},
  booktitle = {Proceedings of the ACM Workshop on Multimedia and Security},
  pages     = {27--30},
  year      = {2001},
  doi       = {10.1145/1232454.1232469}
}

@article{dumitrescu2003,
  author    = {Sorin Dumitrescu and Xiaolin Wu and Zhe Wang},
  title     = {Detection of {LSB} steganography via sample pair analysis},
  journal   = {IEEE Transactions on Signal Processing},
  volume    = {51},
  number    = {7},
  pages     = {1995--2007},
  year      = {2003},
  doi       = {10.1109/TSP.2003.812753}
}

@article{pevny2010,
  author    = {Tom{\'a}{\v{s}} Pevn{\'y} and Tom{\'a}{\v{s}} Filler and Patrick Bas},
  title     = {Using high-dimensional image models to perform highly undetectable steganography},
  journal   = {Information Hiding},
  volume    = {6387},
  pages     = {161--177},
  year      = {2010}
}

@article{filler2011,
  author    = {Tom{\'a}{\v{s}} Filler and Jan Judas and Jessica Fridrich},
  title     = {Minimizing additive distortion in steganography using syndrome-trellis codes},
  journal   = {IEEE Transactions on Information Forensics and Security},
  volume    = {6},
  number    = {3},
  pages     = {920--935},
  year      = {2011},
  doi       = {10.1109/TIFS.2011.2134094}
}

@inproceedings{holub2012,
  author    = {Vojt{\v{e}}ch Holub and Jessica Fridrich},
  title     = {Designing steganographic distortion using directional filters},
  booktitle = {Proceedings of IEEE International Workshop on Information Forensics and Security},
  pages     = {234--239},
  year      = {2012},
  doi       = {10.1109/WIFS.2012.6412655}
}

@article{holub2014,
  author    = {Vojt{\v{e}}ch Holub and Jessica Fridrich and Tom{\'a}{\v{s}} Denemark},
  title     = {Universal distortion function for steganography in an arbitrary domain},
  journal   = {EURASIP Journal on Information Security},
  volume    = {2014},
  number    = {1},
  pages     = {1},
  year      = {2014},
  doi       = {10.1186/1687-417X-2014-1}
}

@article{fridrich2012rich,
  author    = {Jessica Fridrich and Jan Kodovsk{\'y}},
  title     = {Rich models for steganalysis of digital images},
  journal   = {IEEE Transactions on Information Forensics and Security},
  volume    = {7},
  number    = {3},
  pages     = {868--882},
  year      = {2012},
  doi       = {10.1109/TIFS.2012.2190402}
}

@article{kodovsky2012ensemble,
  author    = {Jan Kodovsk{\'y} and Jessica Fridrich and Vojt{\v{e}}ch Holub},
  title     = {Ensemble classifiers for steganalysis of digital media},
  journal   = {IEEE Transactions on Information Forensics and Security},
  volume    = {7},
  number    = {2},
  pages     = {432--444},
  year      = {2012},
  doi       = {10.1109/TIFS.2011.2175919}
}

@article{yedroudj2018,
  author    = {Mounir Yedroudj and Fr{\'e}d{\'e}ric Comby and Marc Chaumont},
  title     = {{Yedroudj-Net}: An efficient {CNN} for spatial steganalysis},
  journal   = {Proceedings of ICASSP},
  pages     = {2092--2096},
  year      = {2018},
  doi       = {10.1109/ICASSP.2018.8461438}
}

@article{boroumand2019,
  author    = {Masoud Boroumand and Mo Chen and Jessica Fridrich},
  title     = {Deep residual network for steganalysis of digital images},
  journal   = {IEEE Transactions on Information Forensics and Security},
  volume    = {14},
  number    = {5},
  pages     = {1181--1193},
  year      = {2019},
  doi       = {10.1109/TIFS.2018.2871749}
}

@article{khurshid2020,
  author    = {Farhana Khurshid and A. H. Mir},
  title     = {Fuzzy logic-based adaptive image steganography for secure digital communication},
  journal   = {Multimedia Tools and Applications},
  volume    = {79},
  pages     = {26777--26798},
  year      = {2020},
  doi       = {10.1007/s11042-020-09203-z}
}

@inproceedings{abdulla2014,
  author    = {A. A. Abdulla and H. Sellahewa and S. A. Jassim},
  title     = {Steganography based on pixel intensity value decomposition},
  booktitle = {SPIE Mobile Multimedia/Image Processing, Security, and Applications},
  year      = {2014}
}

@article{hussain2018,
  author    = {Mehdi Hussain and Ainuddin Wahid Abdul Wahab and Yamani Idna Bin Idris and Anthony T. S. Ho and Ki-Hyun Jung},
  title     = {Image steganography in spatial domain: A survey},
  journal   = {Signal Processing: Image Communication},
  volume    = {65},
  pages     = {46--66},
  year      = {2018},
  doi       = {10.1016/j.image.2018.03.012}
}

@article{systematic2024,
  author    = {A. B. Ogundokun and O. A. Arulogun and others},
  title     = {Image steganography techniques for resisting statistical steganalysis attacks: A systematic literature review},
  journal   = {PLOS ONE},
  volume    = {19},
  number    = {9},
  pages     = {e0308807},
  year      = {2024}
}

@article{zadeh1965,
  author    = {Lotfi A. Zadeh},
  title     = {Fuzzy sets},
  journal   = {Information and Control},
  volume    = {8},
  number    = {3},
  pages     = {338--353},
  year      = {1965},
  doi       = {10.1016/S0019-9958(65)90241-X}
}

@article{mamdani1975,
  author    = {Ebrahim H. Mamdani and Sedrak Assilian},
  title     = {An experiment in linguistic synthesis with a fuzzy logic controller},
  journal   = {International Journal of Man-Machine Studies},
  volume    = {7},
  number    = {1},
  pages     = {1--13},
  year      = {1975},
  doi       = {10.1016/S0020-7373(75)80002-2}
}

@book{ross2010,
  author    = {Timothy J. Ross},
  title     = {Fuzzy Logic with Engineering Applications},
  publisher = {Wiley},
  edition   = {3},
  year      = {2010}
}

@article{wang1992,
  author    = {Li-Xin Wang and Jerry M. Mendel},
  title     = {Generating fuzzy rules by learning from examples},
  journal   = {IEEE Transactions on Systems, Man, and Cybernetics},
  volume    = {22},
  number    = {6},
  pages     = {1414--1427},
  year      = {1992},
  doi       = {10.1109/21.199466}
}

@article{shannon1948,
  author    = {Claude E. Shannon},
  title     = {A mathematical theory of communication},
  journal   = {Bell System Technical Journal},
  volume    = {27},
  number    = {3},
  pages     = {379--423},
  year      = {1948},
  doi       = {10.1002/j.1538-7305.1948.tb01338.x}
}

@article{wang2004,
  author    = {Zhou Wang and Alan C. Bovik and Hamid R. Sheikh and Eero P. Simoncelli},
  title     = {Image quality assessment: From error visibility to structural similarity},
  journal   = {IEEE Transactions on Image Processing},
  volume    = {13},
  number    = {4},
  pages     = {600--612},
  year      = {2004},
  doi       = {10.1109/TIP.2003.819861}
}

@misc{rfc9106,
  author       = {{IRTF CFRG}},
  title        = {{RFC 9106}: Argon2 Memory-Hard Function for Password Hashing and Proof-of-Work Applications},
  year         = {2021},
  howpublished = {\url{https://www.rfc-editor.org/rfc/rfc9106}}
}

@misc{nistgcm,
  author       = {{National Institute of Standards and Technology}},
  title        = {{NIST SP 800-38D}: Recommendation for Block Cipher Modes of Operation: Galois/Counter Mode (GCM) and GMAC},
  year         = {2007},
  howpublished = {\url{https://csrc.nist.gov/publications/detail/sp/800-38d/final}}
}

@article{zhang2011,
  author    = {Xinpeng Zhang},
  title     = {Reversible data hiding in encrypted images},
  journal   = {IEEE Signal Processing Letters},
  volume    = {18},
  number    = {4},
  pages     = {255--258},
  year      = {2011},
  doi       = {10.1109/LSP.2011.2114651}
}

@inproceedings{filler2010,
  author    = {Tom{\'a}{\v{s}} Filler and Jan Judas and Jessica Fridrich},
  title     = {Minimizing embedding impact in steganography using trellis-coded quantization},
  booktitle = {SPIE Media Forensics and Security},
  volume    = {7541},
  year      = {2010}
}

@inproceedings{bas2011boss,
  author    = {Patrick Bas and Tom{\'a}{\v{s}} Filler and Tom{\'a}{\v{s}} Pevn{\'y}},
  title     = {Break our steganographic system: The ins and outs of organizing {BOSS}},
  booktitle = {Information Hiding},
  pages     = {59--70},
  publisher = {Springer},
  year      = {2011}
}

@misc{bows2007,
  author       = {Patrick Bas and Teddy Furon},
  title        = {{BOWS-2} contest},
  year         = {2007},
  howpublished = {\url{http://bows2.ec-lille.fr}}
}

@article{ker2013,
  author    = {Andrew D. Ker and Tom{\'a}{\v{s}} Pevn{\'y} and Jessica Fridrich and Patrick Bas},
  title     = {Moving steganography and steganalysis from the laboratory into the real world},
  journal   = {Proceedings of the 1st ACM Workshop on Information Hiding and Multimedia Security},
  pages     = {45--58},
  year      = {2013}
}

@article{denemark2015,
  author    = {Tom{\'a}{\v{s}} Denemark and Jan Kodovsk{\'y} and Jessica Fridrich},
  title     = {Steganalysis features for content-adaptive {JPEG} steganography},
  journal   = {IEEE Transactions on Information Forensics and Security},
  volume    = {10},
  number    = {8},
  pages     = {1736--1746},
  year      = {2015}
}

@article{alrusaini2025,
  author    = {O. A. Alrusaini and others},
  title     = {Deep learning for steganalysis: Evaluating model robustness under image transformations},
  journal   = {Frontiers in Artificial Intelligence},
  volume    = {8},
  pages     = {1532895},
  year      = {2025}
}

\end{document}